\documentclass[sn-vancouver]{sn-jnl}

\usepackage{natbib}
\usepackage{booktabs}
\usepackage{makecell}
\usepackage{bm} 
\usepackage{hhline}
\usepackage{colortbl}
\usepackage{multicol}
\usepackage{subcaption}
\usepackage[dvipsnames]{xcolor}
\UseRawInputEncoding
\jyear{2022}%

\usepackage{verbatim}

\theoremstyle{thmstyleone}%
\theoremstyle{thmstyletwo}%

\theoremstyle{thmstylethree}%

\begin{document}
\title[Article Title]{A generalized framework to predict continuous scores from medical ordinal labels}


\author[1,2]{\fnm{Katharina V.} \sur{Hoebel}}\email{khoebel@mit.edu}
\equalcont{These authors contributed equally to this work.}

\author[1,3]{\fnm{Andreanne}
\sur{Lemay}}\email{andreanne.lemay@polymtl.ca}
\equalcont{These authors contributed equally to this work.}

\author[4]{\fnm{John Peter} \sur{Campbell}}\email{campbelp@ohsu.edu}

\author[4]{\fnm{Susan} \sur{Ostmo}}\email{ostmo@ohsu.edu}

\author[5]{\fnm{Michael F.} \sur{Chiang}}\email{michael.chiang@nih.gov}

\author[1,6]{\fnm{Christopher P.} \sur{Bridge}}\email{cbridge@partners.org}

\author[1]{\fnm{Matthew D.} \sur{Li}}\email{MDLI@mgh.harvard.edu}

\author[1]{\fnm{Praveer} \sur{Singh}}\email{psingh19@mgh.harvard.edu}

\author[4]{\fnm{Aaron S.} \sur{Coyner}}\email{coyner@ohsu.edu}


\author*[1,7]{\fnm{Jayashree} \sur{Kalpathy-Cramer}}\email{jkalpathy-cramer@mgh.harvard.edu}

\affil[1]{\orgdiv{Martinos Center for Biomedical Imaging}, \orgaddress{\city{Boston}, \state{MA}, \country{USA}}}

\affil[2]{\orgdiv{Massachusetts Institute of Technology}, \orgaddress{\city{Cambridge}, \state{MA}, \country{USA}}}

\affil[3]{\orgdiv{NeuroPoly}, \orgname{Polytechnique Montreal}, \orgaddress{\city{Montreal}, \state{QC}, \country{Canada}}}

\affil[4]{\orgname{Oregon Health and Science University}, \orgaddress{\city{Portland}, \state{OR}, \country{USA}}}

\affil[5]{\orgname{National Eye Institute}, \orgaddress{\city{Bethesda}, \state{MD}, \country{USA}}}

\affil[6]{\orgdiv{MGH \& BWH Center for Clinical Data Science}, \orgaddress{\city{Boston}, \state{MA}, \country{USA}}}


\affil[7]{\orgname{University of Colorado Anschutz Medical Campus}, \orgaddress{\city{Aurora}, \state{CO}, \country{USA}}}




\abstract{
Background: Many variables of interest in clinical medicine, like disease severity, are recorded using discrete ordinal categories such as normal/mild/moderate/severe. 
These labels are used to train and evaluate disease severity prediction models.
However, ordinal categories represent a simplification of an underlying continuous severity spectrum.
Using continuous scores instead of ordinal categories is more sensitive to detecting small changes in disease severity over time.  
Here, we present a generalized framework that accurately predicts continuously valued variables using only discrete ordinal labels during model development. 

Methods: We study the framework using three datasets: disease severity prediction for retinopathy of prematurity and knee osteoarthritis and breast density prediction from mammograms.
For each dataset deep learning models are trained using discrete labels, and the model outputs were converted into continuous scores.
The quality of the continuously valued predictions was compared to expert severity scores that were more detailed than the discrete training labels.
We study the performance of conventional and Monte Carlo dropout multi-class classification, ordinal classification, regression, and Siamese models. 

Findings: 
We found that for all three clinical prediction tasks, models that take the ordinal relationship of the training labels into account outperformed conventional multi-class classification models. 
Particularly the continuous scores generated by ordinal classification and regression models showed a significantly higher correlation with expert rankings of disease severity and lower mean squared errors compared to the multi-class classification models. 
Furthermore, the use of MC dropout significantly improved the ability of all evaluated deep learning approaches to predict continuously valued scores that truthfully reflect the underlying continuous target variable. 

Interpretation: 
We showed that accurate continuously valued predictions can be generated even if the model development only involves discrete ordinal labels. The novel framework has been validated on three different clinical prediction tasks and has proven to bridge the gap between discrete ordinal labels and the underlying continuously valued variables.}

\maketitle
\section{Introduction\label{sec1}}
\begin{figure}
     \centering
        \includegraphics[width=1.0\textwidth]{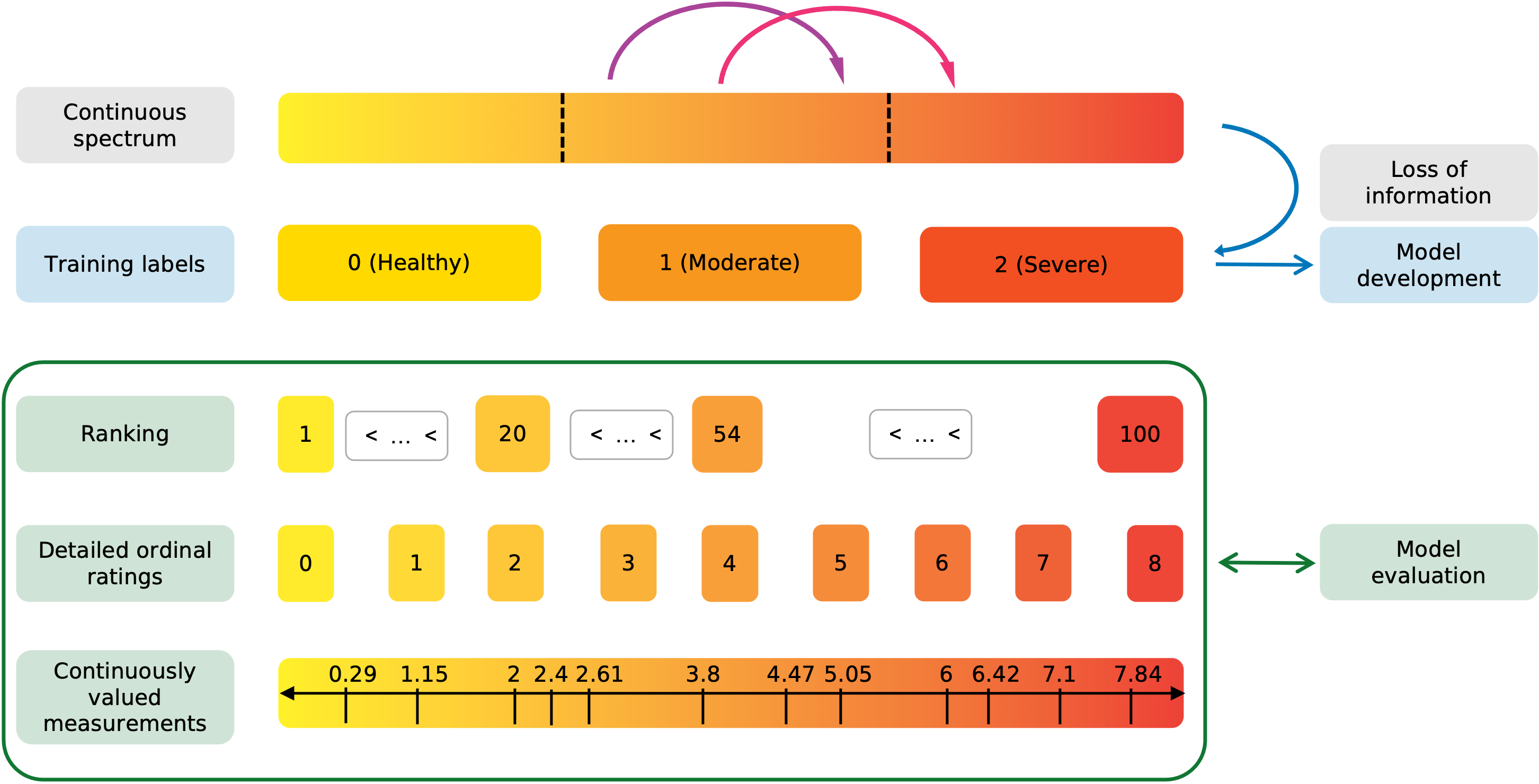}
        \caption{\textbf{Relationship between the underlying continuous variable of interest and the training and evaluation labels.} The conversion of a latent continuously distributed variable into discrete ordinal variables for model development represents a loss of information. The purple and magenta arrows represent temporal changes in disease severity. The available annotation types for evaluating the continuous predictions are presented in the green box: Rankings, ordinal ratings on a more detailed scale than the training labels, and continuously valued measurements. }
        \label{fig:visual_abstract}
\end{figure}

Many clinical variables, like disease severity, are communicated and recorded as discrete ordinal classes. 
However, in reality they are distributed on a continuous spectrum. \citep{Campbell2016PlusVariability}
The discretization of continuously valued variables facilitates documentation and communication and standardizes treatment decisions at the cost of losing information. 
As illustrated in Figure \ref{fig:visual_abstract}, two patients that fall into the same severity category will receive the same label.
Consequently, if their data is used to develop a prediction model, it will be treated exactly the same during training, regardless of their exact position along the continuous spectrum. 
\\
The ability of Deep Learning (DL) to learn and recognize subtle patterns from large amounts of data has led to tremendous successes in automated medical image analysis. \citep{Anwar2018MedicalReview}
DL models have achieved and, in some cases, even exceeded human performance in disease detection and automatic severity classification for numerous diseases such as diabetic retinopathy, retinopathy of prematurity, and osteoarthritis. \citep{Gulshan2016DevelopmentPhotographs, Brown2018AutomatedNetworks,  Tiulpin2018AutomaticApproach} 
\\
Yet, these successes have been build on the simplifying assumption that disease severity prediction can be formulated as a simple classification task. 
\\
Researchers mostly use DL architectures that are intended for the classification of nominal categories and ignore the inherent ordinal nature of the available training labels. 
More so, disease severity prediction tasks are often simplified even more by treating them as binary problems, e.g., the identification of severe disease \citep{Stidham2019PerformanceColitis}, cases for referral \citep{DeFauw2018ClinicallyDisease, Leibig2017LeveragingDetection}, or disease detection \citep{Gulshan2016DevelopmentPhotographs}.



\paragraph{Advantages of continuous scores}
In addition to the class, the position of a case on the continuous spectrum contains valuable clinical information that is not captured by current approaches. \cite{Antony2016QuantifyingNetworks}
Therefore, the use of continuous scores to describe clinical variables that are distributed on a continuous spectrum provides several advantages over discrete ordinal variables. \\
First, continuous metrics allow for the detection and quantification of changes within a class, i.e., an increase in disease severity that does not constitute a transition between class $n$ and $n+1$. \cite{Li2020SiameseImaging,Brown2018FullyLearning}
In Figure \ref{fig:visual_abstract} the purple and magenta arrows represent similar large increases in disease severity.
While the magenta transition would get detected by the traditional classification approach, the purple transition would not. 
The only difference between the two transitions is that the magenta one crosses the class boundary from moderate to severe, while the purple arrow represents a within class change.
The detection of within-class changes allows to detect disease deterioration earlier and act upon it if required. 
\\
Second, the higher degree of information presented in continuous vs. ordinal scores can be useful for efficient patient stratification, particularly the identification of cases close to a decision boundary.
Third, expert perception of class boundaries can be subject to changes over time. \citep{Moleta2017Plus2007}
Therefore, models ignoring the continuous nature of disease severity could become less valuable over time as, e.g., the perception of what constitutes mild versus moderate disease severity shifts.
Lastly, an algorithm that predicts a continuous score is more likely to fulfill notions of individual fairness as similar individuals that are close to the label decision boundaries will be more likely to receive similar scores compared to using a simple classification algorithm. \cite{Dwork2012FairnessAwareness}

\paragraph{Related work} 
Previous attempts to predict  continuous disease severity scores  have made use of either conventional classification networks or Siamese networks.
Redd et al. proposed to aggregate the softmax outputs from a conventional 3-class convolutional network into one continuously valued vascular severity score for disease severity classification in retinopathy of prematurity (ROP). \cite{Redd2019EvaluationPrematurity}
This score strongly correlates with experts' ranking of overall disease severity.
Furthermore, changes in this score over time accurately reflect disease progression. \cite{Brown2018FullyLearning}
However, the training objective of multi-class classification models is to separate the latent space representation of classes as much as possible. This could therefore lead to unstable predictions and confusion at the class boundaries. 
\\
Using Siamese networks, Li et al. showed that the continuously valued difference relative to a reference pool of images correlates with expert's rankings of disease severity and reflects temporal changes in severity in knee osteoarthritis and ROP. \cite{Li2020SiameseImaging}
Similarly, in a study on an x-ray based severity score for COVID-19, a score generated by a Siamese network highly correlated with radiologist-determined severity scores.
However, the performance of Siamese networks for the predictions of continuous scores has not been compared to other methods and their calibration has not been studied yet.
\\

\paragraph{Study aim}
Here, we aim to identify model development strategies that lead to the prediction of accurate continuous scores. 
Most importantly, while we utilize widely available discrete ordinal labels for training, the  models' performance to predict accurate continuous scores is evaluated using labels on a finer scale than the training ground truth (illustrated in the green box in Figure \ref{fig:visual_abstract}) on three datasets: disease severity prediction for ROP and knee osteoarthritis and breast density estimation from mammograms. 
Following this process, we aim to show that it is possible to develop models that are capable to recover the information lost through the discretization of the continuous target variable.

\paragraph{}

\section{Methods}\label{sec11}

\subsection{Datasets}
All images were de-identified prior to data access; ethical approval for this study was therefore not required.
Dataset splits were performed on a patient level. The size of all dasets is listed in Table \ref{tab:dataset} and class distributions for each dataset are listed in Appendix \ref{app:data}.

\subsubsection{ROP}
ROP is an eye disorder mainly developed by prematurely born babies and is among the leading causes of preventable childhood blindness. \cite{shah2016retinopathy} 
It is characterized by a continuous spectrum of abnormal growth of retinal blood vessels which is typically categorized into three discrete severity classes: normal, pre-plus, or plus. \citep{Quinn2005ThePrematurity, Campbell2016PlusVariability} \\
We use the same dataset, labels, and preprocessing as described by Brown et al.\citep{Brown2018AutomatedNetworks}
In addition to the standard diagnostic labels, the test set was labeled by five raters on a scale from 1 to 9 and five experts ranked an additional 100 ROP photographs based on severity.\citep{Taylor2019MonitoringLearning, Brown2018AutomatedNetworks}

\subsubsection{Knee osteoarthritis}
The global prevalence of knee osteoarthritis is 22.9\% for individuals over 40, causing chronic pain and functional disability. \cite{cui2020global} 
Knee osteoarthritis can be diagnosed with radiographic images and disease severity is typically evaluated using the Kellgren-Lawrence (KL) scale consisting of the following severity categories: none, doubtful, mild, moderate, and severe. \citep{kellgren1957radiological}\\
We use the the Multicenter Osteoarthritis Study (MOST) dataset. 100 images from the test were ranked by their severity by three experts. \citep{Li2020SiameseImaging}
All images were center cropped to 224x224 pixels and intensity scaled between 0 and 1 as preprocessing.  

\subsubsection{Breast density}
Breast density is typically categorized as fatty, scattered, heterogeneous, or dense, depending on the amount of fibroglandular tissue present.\citep{Liberman2002BreastBI-RADS} 
Women with high breast density are at a higher risk of developing breast cancer and require additional MRI screening. \citep{Boyd1995QuantitativeStudy, Bakker2019SupplementalTissue}\\ 
We use a subset of the Digital Mammographic Imaging Screening Trial (DMIST) dataset. \citep{Pisano2005DiagnosticScreening} 
Furthermore, for 1892 mammographs from the test dataset, an automatic assessment of the volumetric breast density was obtained using the commercially available Volpara Density software which has demonstrated a good agreement with expert ratings of breast density  (see Figure \ref{fig:breast_density}). \citep{highnam2010robust, wanders2017volumetric}
Preprocessed mammograms were of size 224x224 pixels.

\begin{table}[hbtp]
\centering
{\caption{Summary of dataset size and training/validation/test splits of the three datasets used for this study: disease severity prediction in retinopathy of prematurity (ROP) and knee osteoarthritis (OA) and breast density prediction} \label{tab:dataset}}
{\begin{tabular}{lcccc}
  \toprule
  \textbf{Dataset} & \textbf{Size} & \textbf{Training} & \textbf{Validation}  & \textbf{Test} \\
  \midrule
   ROP & 5611 &  4322& 722  & 467 (9-point scale)\\
   &&&& 100 (ranked)\\
  \midrule
  Knee OA & 14273 & 12268&1905&100 (ranked)\\
  \midrule
  Breast Density & 83034 & 70293 & 10849 & 1892 (Volpara Density)\\
  \bottomrule
  \end{tabular}}
\end{table}

\subsection{Model training}

\subsection{Model types}
\label{sec:model_types}
Four model types were trained: multi-class and ordinal classification, regression, and Siamese. The model output was converted to a continuous score value for each model to represent the underlying severity spectrum of the medical tasks studies. 

\paragraph{Classification}
All classification models for this study were trained with cross-entropy loss. 
The continuous severity score is computed as the of the softmax outputs weighted by their class (Equation \ref{eq:multiclass}), leading to scores from 0 to k-1.
 
\begin{equation} \label{eq:multiclass}
Cl_{score} = \sum_{i=1}^{k} p_{i} \times i - 1
\end{equation}
 
\noindent with $k$ being the number of classes and $p_{i}$ the softmax probability of class $i$.
 
\paragraph{Ordinal classification}
In ordinal classification, the task is broken up into a $k - 1$ binary classification tasks, leading to one output unit less than the number of classes. \citep{Li2006OrdinalClassification}
During training, the ordinal loss function penalizes larger misclassification errors more than smaller errors (e.g., predicting class 2 when the ground truth label is 0 is penalized more than if the models predicts class 1). We use the CORAL loss as described by Cao et al. for model optimization.\cite{Cao2020RankEstimation}

 


A continuous score is generated by summing over the output probabilities (Equation \ref{eq:ordinal}), resulting in values ranging from 0 to k-1.

\begin{equation} \label{eq:ordinal}
O_{score} = \sum_{i=1}^{k} p_{i}
\end{equation}
 

\paragraph{Regression}
Similar to ordinal models, regression models require the ordinality of the target output. However, unlike ordinal models, the output of regression models is a continuous value rather than a discrete class. 
The regression models were trained using the mean squared error loss function with the class number as the target value. The raw model output yields a continuous value; hence, no conversion is required to receive a continuous score.   

\paragraph{Siamese}
Siamese models compare pairs of images to evaluate their similarity.
They are composed of two branches consisting of identical sub-networks with shared weights where each of the two images is processed in one of the branches.
The lower the Euclidean distance between the outputs of each branch the higher the similarity between the inputs.
Following a procedure described by Li et al., at test time, the target images are compared to ten anchor images associated with class 0. \cite{Li2020SiameseImaging}
Here, the continuous score is the median of the Euclidean distances between the target and the ten anchor images.

 

\subsection{Monte Carlo dropout}
By utilizing dropout not just during training but also at test time, yields $N$ slightly different Monte Carlo (MC) predictions.\citep{gal2016dropout}
The MC predictions can subsequently be averaged, to obtain the final output prediction. 
All models referred to as \textit{MC models} were trained with spatial dropout after each residual block of the ResNets models, and $N=50$ MC iterations at test time.
The dropout rates are $0{\cdot }2$, $0{\cdot }2$, and $0{\cdot }1$ for ROP, knee osteoarthritis, and breast density, respectively, and were selected empirically and based on current literature. 

\subsection{Model training}
Model parameters were selected based on initial data exploration and empirical results. All ROP models were using a ResNet18 and all knee osteorarthritis and breast density models a ResNet50. A detailed description of the training parameters can be found in Appendix \ref{app:training}.

\subsection{Evaluation}
\subsubsection{Metrics}
\paragraph{Ranked datasets}
The model performance was evaluated based on the ranked test data using the following three metrics. 
First, we computed Spearman's rank coefficient between the rank and the predicted score. 
A monotonic increase between both metrics is expected; hence, a Spearman coefficient of 1 corresponds to a perfect correlation.
Second, we computed agreement between the ground truth rank and the rank based on the continuous score using mean squared error (MSE) to quantify the correspondence between the predictions and ground truth. 
Here, the ranks were normalized to the maximum rank.
Finally, the classification performance was assessed using clinically relevant AUCs. 
We defined the clinically relevant classification and normal/pre-plus vs. plus for ROP, none/doubtful vs. mild/moderate/severe for knee osteoarthritis, and fatty/scattered vs. heterogeneous/dense for breast density. 

\paragraph{ROP}
A subset of the ROP test set had expert ratings from 1 to 9 based on the quantitative scale previously published by Taylor et al. \citep{Taylor2019MonitoringLearning} 
The correspondence between the expert rating and the continuous predicted scores were measured using the MSE.  

\subsubsection{Statistical analysis}
Metrics were bootstrapped (500 iterations) and 95\% confidence intervals were evaluated for statistical analysis. Bootstrapped metrics yielding two-sided $t$-test with a p-value inferior to 5\% were considered statistically different.

\section{Results }
\subsection{Predicted score compared with severity rankings}\label{sec:rank}
\begin{figure}
     \centering
     \begin{subfigure}[b]{\textwidth}
         \centering
         \includegraphics[width=0.9\textwidth]{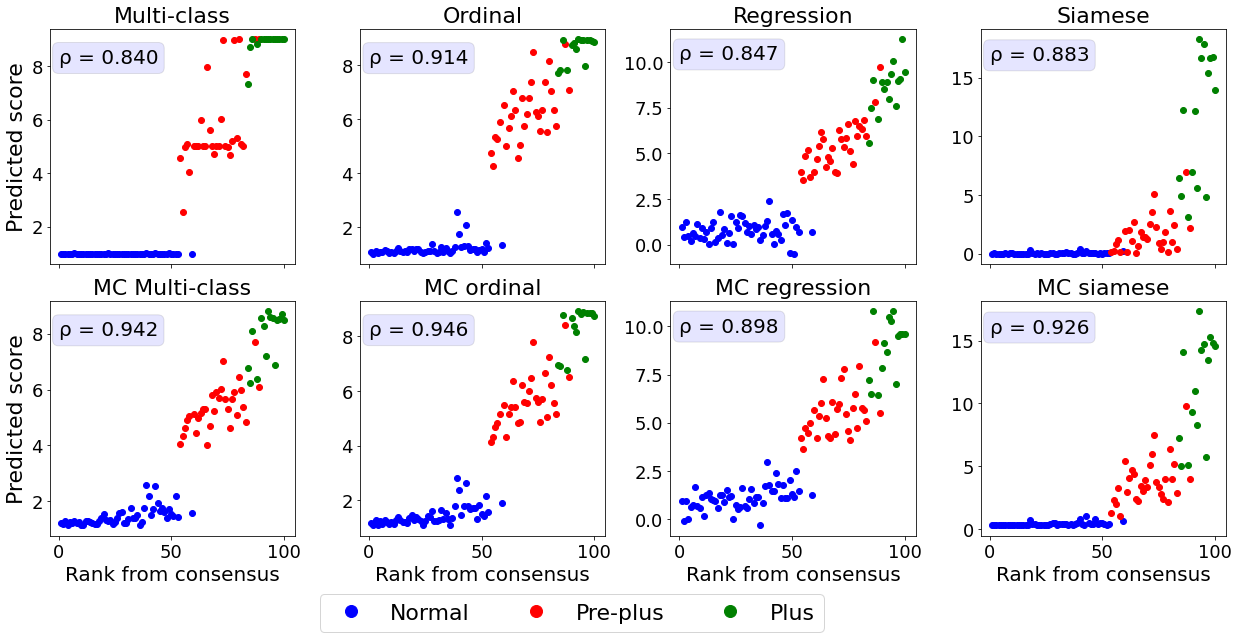}
         \caption{ROP}
         \label{fig:rop_score}
     \end{subfigure}
     \hfill
     \begin{subfigure}[b]{\textwidth}
         \centering
         \includegraphics[width=0.9\textwidth]{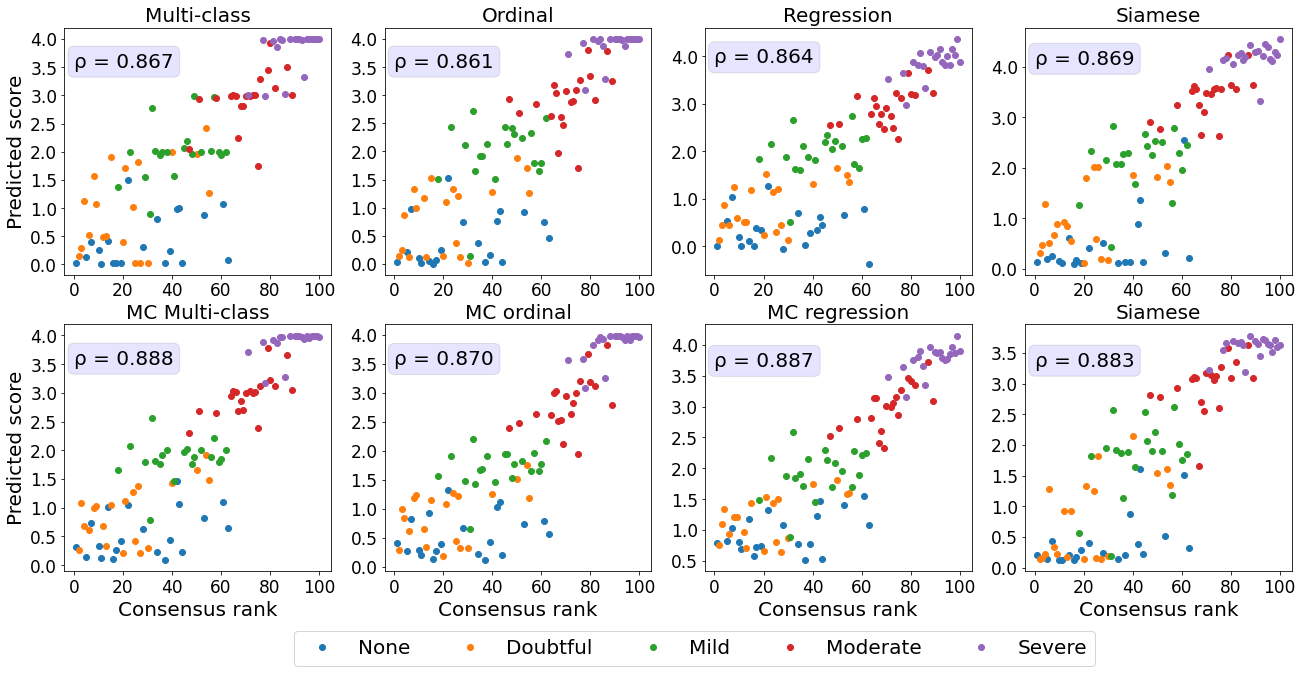}
         \caption{Knee osteoarthritis}
         \label{fig:knee_score}
     \end{subfigure}
     \hfill
     \begin{subfigure}[b]{\textwidth}
         \centering
         \includegraphics[width=0.9\textwidth]{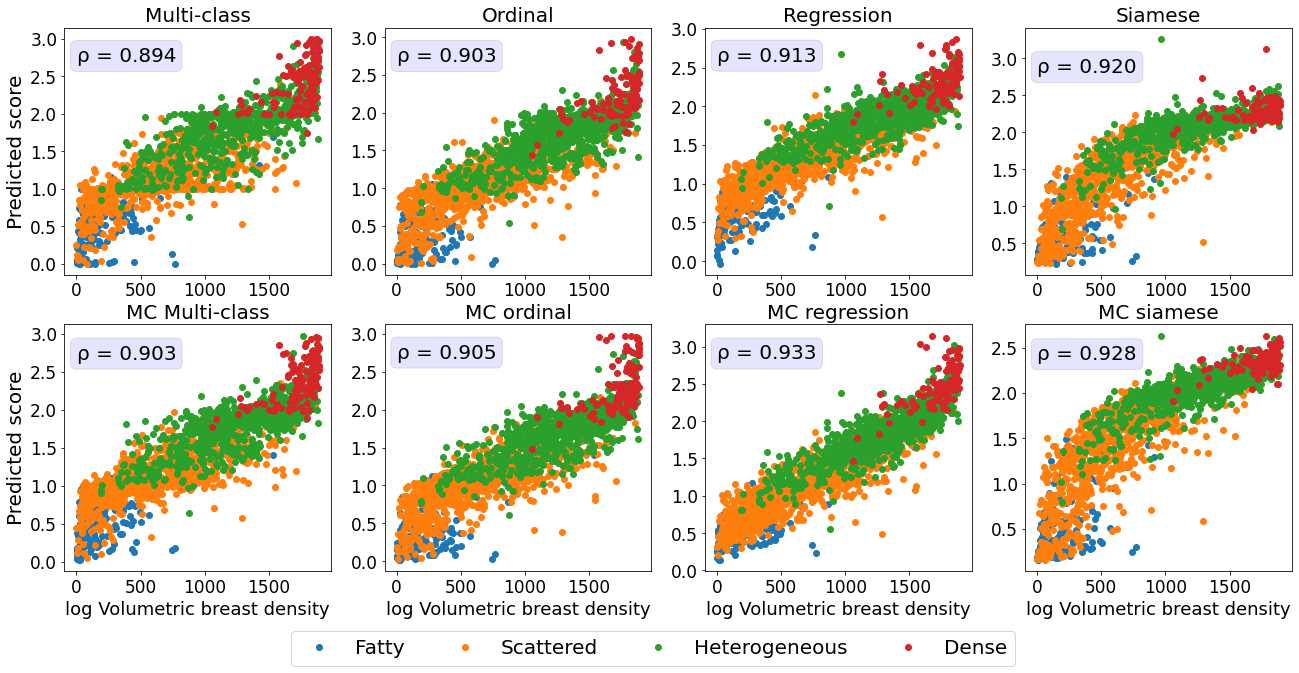}
         \caption{Breast density}
         \label{fig:mammo_score}
     \end{subfigure}
        \caption{\textbf{Correspondence between model predicted score and severity ranking.} For each model, the Spearman correlation coefficient ($\rho$) is displayed in the upper left corner. It indicates the monotonicity of the correlation where 1 is a perfectly increasing correlation and -1 is a perfectly decreasing correlation.}
        \label{fig:rank_vs_pred}
\end{figure}

\paragraph{Agreement between predicted score and severity rankings}
We first assessed how well the predicted continuous scores reflect a ranking of the images in each dataset. 
Retinal photographs and knee radiographs were ranked by domain experts with increasing disease severity. 
The mammograms were ranked with increasing density based on the quantitative continuously valued Volpara breast density score.
\\
The relationship between the ground truth rankings and the predicted continuous scores are presented in Figure \ref{fig:rank_vs_pred}.
For all datasets, the multi-class models without MC dropout display horizontal plateaus around the class boundaries where the predicted score is more or less constant with increasing rank. 
Similar patterns can be observed for the Siamese and MC Siamese models, especially for normal ROP and knee osteoarthritis cases. 

\begin{table}[hbtp]
  
    \centering
  
  {\caption{\small{\textbf{Model performance overview (MEAN $\pm$ 95\% CI)}. Bold values indicate a statistical difference (p-value $ < 0.05$) was observed. Spearman's rank correlation coefficient and the AUC are measured on the predicted continuous score while the MSE is measured between the normalized ground truth rank and the predicted rank generated from continuous scores. AUC was measured between normal and pre-plus vs. plus for ROP, none and doubtful vs. mild, moderate, severe for knee osteoarthritis, and fatty and scattered vs. heterogeneous and dense for breast density.}}
  \label{tab:metric-table}
  }
  
  {\begin{tabular}{lcccc}
  \toprule
  \textbf{Model} & \textbf{MSE} $\downarrow$ & \textbf{Spearman $\uparrow$}  & \textbf{\thead{clinically relevant\\ AUC}} $\uparrow$ \\
  \midrule
  \multicolumn{4}{l}{\textbf{ROP}} \\
  \midrule
  Multi-class & $ 0.027 \pm 0.009$ & $0.84 \pm 0.07$ & $0.98 \pm 0.02$\\
  MC multi-class & \bm{$0.010 \pm 0.003$} & \bm{$0.94 \pm 0.02$} & \bm{$0.99 \pm 0.01$} \\
  Ordinal  & $0.015 \pm 0.005$ & $0.91 \pm 0.04$ & $0.99 \pm 0.01$\\
  MC ordinal & \bm{$0.009 \pm 0.003$} & \bm{$0.94 \pm 0.02$} & $0.99 \pm 0.02$ \\
  Regression & $0.026 \pm 0.010$ & $0.85 \pm 0.07$ & $0.98 \pm 0.03$ \\
  MC regression  & \bm{$0.017 \pm 0.005$} & \bm{$0.89 \pm 0.05$} & \bm{$0.98 \pm 0.02$}\\
  Siamese & $0.020 \pm 0.007$ & $0.88 \pm 0.05$ & \bm{$0.99 \pm 0.01$} \\
  MC Siamese & \bm{$0.013 \pm 0.004$} & \bm{$0.92 \pm 0.03$ }& $0.98 \pm 0.02$  \\
  \midrule
  \multicolumn{4}{l}{\textbf{Knee osteoarthritis}} \\
  \midrule
  Multi-class &  $0.023 \pm 0.008$ & $0.86 \pm 0.06$ &  $0.97 \pm 0.02$ \\
  MC multi-class &  \bm{$0.019 \pm 0.006$} &  \bm{$0.89 \pm 0.05$} & \bm{ $0.99 \pm 0.01$} \\
  Ordinal &  $0.024 \pm 0.007$ &  $0.85 \pm 0.07$ &  $0.98 \pm 0.02$ \\
  MC ordinal &  \bm{$0.022 \pm 0.007$} &  \bm{$0.86 \pm 0.06$} &  \bm{$0.99 \pm 0.02$} \\
  Regression &  $0.023 \pm 0.009$ & $0.86 \pm 0.06$ &  \bm{$0.99 \pm 0.01$} \\
  MC regression &  \bm{$0.019 \pm 0.006$} &  \bm{$0.88 \pm 0.05$} &  $0.98 \pm 0.02$ \\
  Siamese & $0.022 \pm 0.007$ &  $0.87 \pm 0.05$ & \bm{$0.97 \pm 0.03$}  \\
  MC Siamese & \bm{$0.020 \pm 0.005$} & \bm{$0.88 \pm 0.04$} & $0.97 \pm 0.03$ \\
\midrule
  \multicolumn{4}{l}{\textbf{Breast density}} \\
  \midrule
  Multi-class & $0.018 \pm 0.001$ & $0.89 \pm 0.01$ & $0.93 \pm 0.01$ \\
  MC multi-class & \bm{$0.016 \pm 0.001$} & \bm{$0.90 \pm 0.01$} & \bm{$0.94 \pm 0.01$} \\
  Ordinal & $0.016 \pm 0.001$ & $0.90 \pm 0.01$ & $0.93 \pm 0.01$ \\
  MC ordinal & \bm{$0.015 \pm 0.001$} &\bm{$0.91 \pm 0.01$} & \bm{$0.94 \pm 0.01$}\\
  Regression & $0.015 \pm 0.001$ & $0.91 \pm 0.01$ & $0.94 \pm 0.01$ \\
  MC regression & \bm{$0.011 \pm 0.001$} & \bm{$0.93 \pm 0.01$} & $0.94 \pm 0.01$ \\
  Siamese & $0.013 \pm 0.001$ & $0.92 \pm 0.01$ & $0.91 \pm 0.01$ \\
  MC Siamese & \bm{$0.012 \pm 0.001$} &\bm{$0.93 \pm 0.01$} & \bm{$0.92 \pm 0.01$} \\
  
  \bottomrule
  \end{tabular}}
\end{table}

\paragraph{Agreement between predicted and ground truth rankings}
A linear correlation between the predicted continuous score and the consensus rank cannot be assumed as the predicted score will increment variably depending on the severity increase from a patient of rank $n$ and $n + 1$. 
Therefore, we used the Spearman correlation coefficient and MSE to quantify the agreement between the ground truth ranking and the ranking based on the predictions (see Table \ref{tab:metric-table} and Appendix \ref{sec:app_rank}). 
\\
All MC dropout models were associated with a statistically significant higher Spearman correlation coefficient and lower MSE compared to their non-MC counterparts (p-value $< 2{\cdot }2e-4$, see Appendix \ref{sec:stats} for pair-wise statistical comparisons between the models).
The higher Spearman correlation coefficients and lower MSE indicate that the addition of MC dropout during training and inference improves the ability of DL models to correctly rank the images based on the continuous predicitions.
The models with the best correspondence between actual and predicted rank were MC multi-class and MC ordinal models for ROP, MC multi-class and MC regression for knee osteoarthritis, and MC regression and MC Siamese networks for breast density. 

\paragraph{Classification performance}
All MC dropout models showed a slightly higher or comparable classification performance, as assessed by AUC, to their non-MC equivalent (see Table \ref{tab:metric-table}). 
The  only exceptions were the Siamese models for ROP and knee osteoarthritis and the regression knee osteoarthritis model. 
In these three cases, though statistically significant, the AUC of the MC models was only $0{\cdot }01$ (or less) lower the one of their non-MC equivalents. 
The model associated with the best classification did not necessarily correspond to the best continuous severity scores. 
\\
The following models have the overall best performance for each dataset: MC multi-class (knee osteoarthritis), MC ordinal (ROP), and MC regression (breast density).  

\subsection{Comparison of predicted ROP scores with disease severity ratings}
\begin{figure}[htp]
  \centering
    \includegraphics[width=1.0\textwidth]{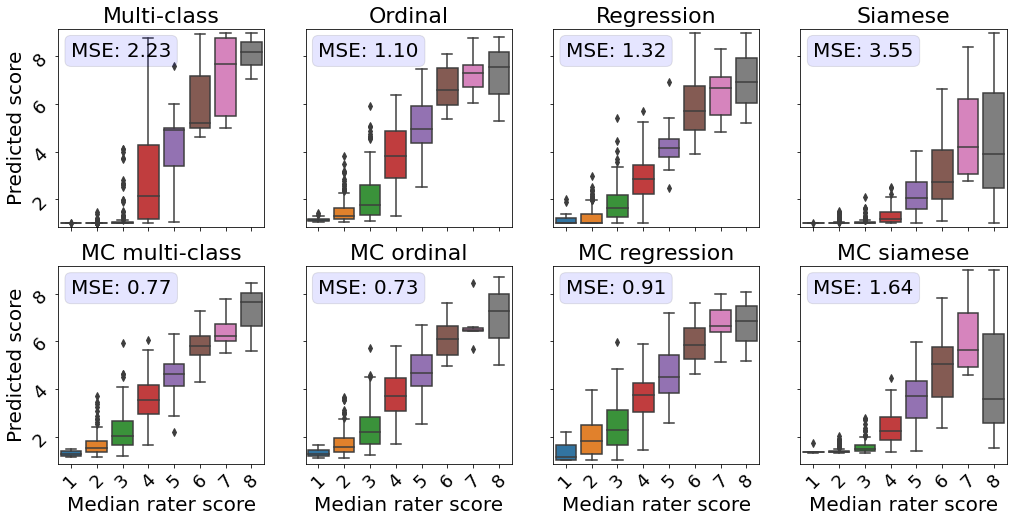}

  \caption{\textbf{Correspondence between predicted and consensus ROP severity on an in-distribution test set.} The consensus ROP score was obtained by calculating the median of five ratings from different experts. The predicted scores from multi-class, ordinal, and regression models that were trained to predict values from 0 to 2 were scaled and shifted to match the 1 to 9 range ($score_{rescaled} = score_{model} \times 2 + 1$). Siamese networks predict values from 0 to infinity and are not fully bounded. The Siamese scores were hence only shifted by 1 ($score_{rescaled} = score_{Siamese} + 1$). In accordance with the severity scale used, Siamese rescaled scores were also clipped to values between 1 and 9. All MSE measurements reported in this figure are statistically different (p-value $< 1.2e-41)$.}
  \label{fig:500val}
\end{figure}

Next, we evaluated the correspondence between the predicted scores and more detailed severity ratings generated by domain experts. 
An subset of the test dataset was rated by five experts on a scale from 1 to 9 instead of the standard scale from 1 to 3. \cite{Taylor2019MonitoringLearning} 
This dataset allowed us to evaluate the quality of the continuous model outputs on a more granular scale than the 3-class labels the models were trained on. 
Perfect continuous predictions would result in increasing disease severity scores with increasing ground severity ratings.
\\
All MC models showed a higher correspondence between the true severity ratings and predicted scores, as reflected by a lower MSE in comparison with their conventional counterparts (see Figure \ref{fig:500val}). 
The models predicting the experts' ratings the most accurately are the MC multi-class and MC ordinal models. 
\\
While Siamese networks showed decent correspondence between the predicted score and the ranked severity , a direct comparison with the severity ratings reveals that the predictions from these models are not well calibrated. 
\\
The multi-class model without MC showed the second worst performance in this analysis. 
Images rated from 1 to 3 by experts mainly obtained scores near 0, which does not highlight the severity differences as perceived by human experts. 
Furthermore, for retinal photographs associated with a score of 4, the model predicted values on the entire spectrum, i.e., from 1 to 9, which is undesirable.   

\begin{figure}[htp]
  \centering
    \includegraphics[width=1.0\textwidth]{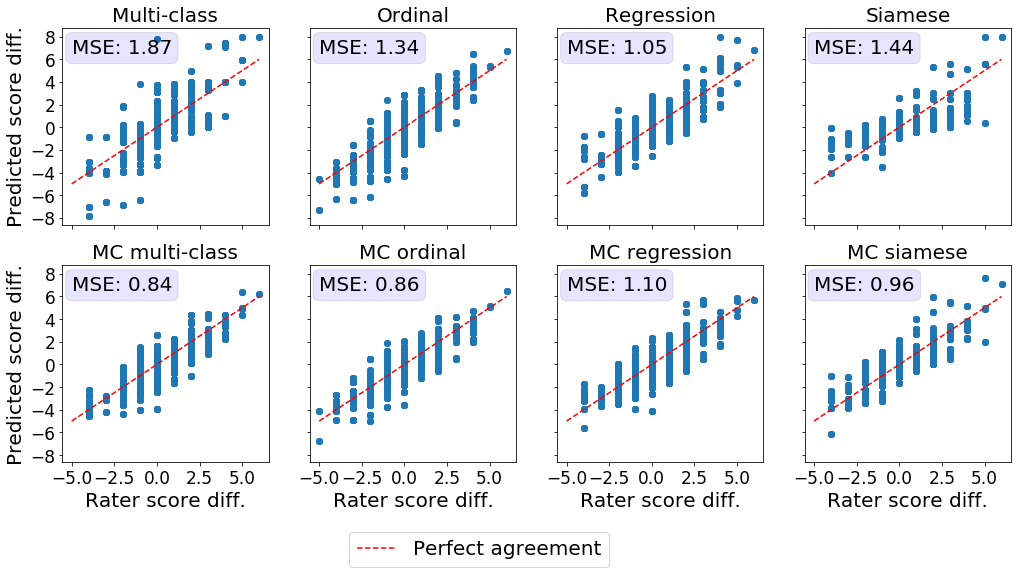}

  \caption{\textbf{Correspondence between the perceived rater score difference on longitudinal images from the same patient and the predicted score difference.} The red dashed line is the identity line and indicates the expected region were the data points should fall. All MSE measurements reported in this figure are statistically different $(p-value < 4{ \cdot }3e-28)$.}
  \label{fig:longitudinal}
\end{figure}

\paragraph{Detection of temporal changes in disease severity}
Another important characteristic of a reliable severity score is its ability to reflect slight changes in disease severity over time. 
The disease evolution was quantified as the difference in the ground truth severity ratings or predicted severity scores between photographs of the same patient taken at different time points.
We then compared the difference in experts' ratings to the difference in the predicted scores using MSE (see Figure \ref{fig:longitudinal}).
Ideally, the difference in the expert's scores should be equal to the difference in the models' predictions. 
MC dropout improved the correspondence between the disease evolution as perceived by experts and predicted by the DL models for multi-class, ordinal, and Siamese models. 
Prediction differences from MC multi-class and MC ordinal models matched the severity shifts in the experts' ratings most closely. 
The conventional multi-class model presents multiple outliers and is associated with the highest MSE.  

\subsection{Comparing predicted breast density scores with continuously valued ground truth breast density measurements}
\begin{figure}[htp]
  \centering
    \includegraphics[width=1.0\textwidth]{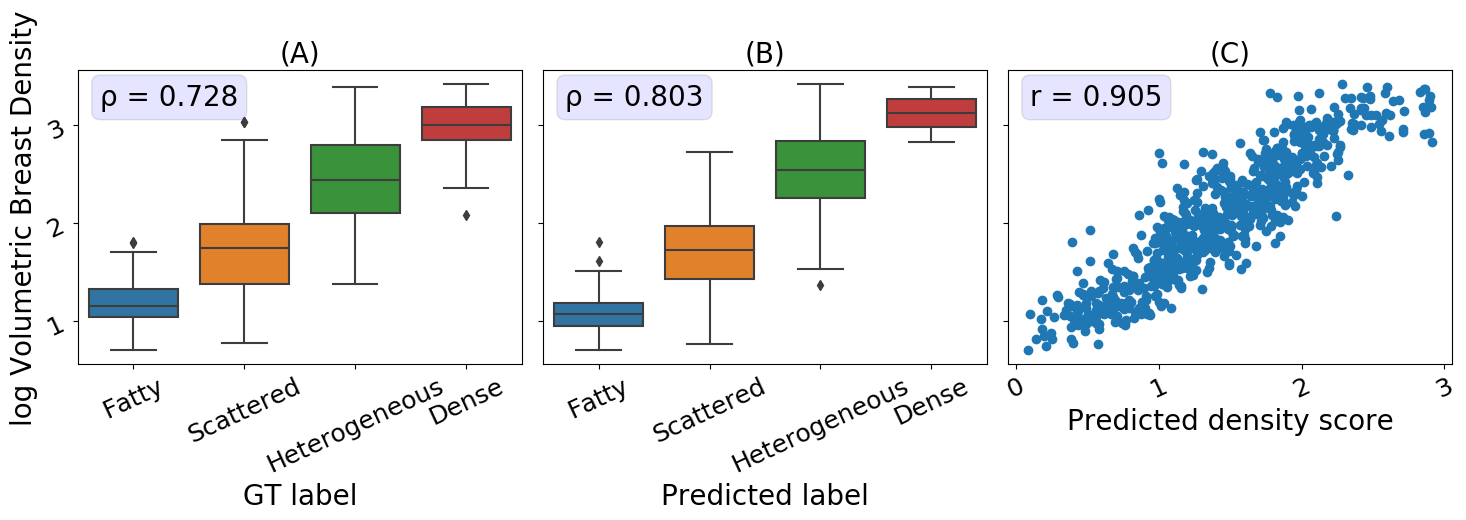}

  \caption{\textbf{Correspondence between volumetric breast density measurement and predicted or true breast density.} The predicted values are from the breast density MC multi-class model. (A) demonstrates the relationship between the expert ratings and the quantitative Volpara measurement. (B) illustrates the correspondence between the predicted labels obtained by taking the class with the highest softmax score and volumetric breast density. (C) plots the continuous predicted score against the volumetric breast density. $\rho$ is the Spearman correlation coefficient for each metric pair.}
  \label{fig:breast_density}
\end{figure}

Lastly, we evaluated the ability of the breast density prediction models trained to accurately reflect the continuously valued Volpara Density measurements. 
The subset of mammograms with the Volpara Densitymeasurements provided us with the unique opportunity to evaluate algorithms trained using ordinal labels on a continuously valued ground truth. 
Therefore, unlike with the ranked score analysis presented in Section \ref{sec:rank}, here we directly compared the Volpara Density scores with the continuously valued model predictions. 
Ideal continuously valued predictions would correlate linearly with the Volpara Density scores. 
\\
We first assessed the relationship between the Volpara Density scores and the discrete ground truth labels generated by domain experts used for training.
As illustrated in the boxplot in Figure \ref{fig:breast_density}A, and by the Spearman correlation coefficient of 0.73, there is a high agreement between the ground truth labels and the Volpara Density scores.
The MC multiclass model's predictions, both class and continuous score, are a close proxy to the volumetric breast density measurements, as seen in Figure \ref{fig:breast_density}B and \ref{fig:breast_density}C with a Spearman correlation coefficient of 0.803 (classification) and Pearson correlation coefficient of 0.91 (continuous scores). 
The high correlation between the  continuous breast density predictions and Volpara Density measurements indicates that our model is able to generate an accurate continuous prediction while being trained on only a finite number of classes.  

\section{Discussion}\label{sec12}
The underlying continuous nature of many prediction targets for DL image analysis tasks, such as breast density and disease severity, has to be taken into account in the process of model design. 
Here, we studied the capability of DL models to intrinsically learn a continuous score while being trained using discrete ordinal labels.
Our results show that training a conventional multi-class classification model without MC dropout does not lead to predictions that reflect the underlying continuous nature of the target variable. 
Approaches that model the relationship between the ordinal labels, such as ordinal classification, regression, and Siamese networks, provide continuous predictions that closely capture the continuity of the target variable even without the use of MC dropout. 
Finally, using MC dropout during training and inference increased the ability of the DL models to predict meaningful continuous scores. 
MC dropout multi-class classification ranked among the best performing models in this study.

\paragraph{Multi-class classification models} 
Ignoring the ordinal relationship between the training label classes causes conventional multi-class prediction models to return predictions that are clustered around the values of the training labels.
This behavior is reflected in the plateaus visible in Figure \ref{fig:rank_vs_pred}, and the medians in Figures \ref{fig:500val}, and \ref{fig:nepal}, a lower Spearman correlation coefficient and higher MSE.
\\
Due to the definition of the training objective, multi-class classification models are optimized to precisely predict a specific class and discouraged from predicting scores at the class boundaries. 
This behavior is desirable for nominal classification, where the classes should be separated as clearly as possible with minimal overlap in the feature latent space to avoid ambiguous predictions.
However, the approach is not appropriate for problems with a target variable with an underlying continuous nature and explains the limited performance of the multi-class classification models to predict meaningful continuous scores.  

\paragraph{Siamese networks}
Siamese networks showed decent correspondence between the ranked severity and the predicted score (Figure \ref{fig:rank_vs_pred}.
However, a direct comparison between the predicted score and the severity determined by domain experts (Figure \ref{fig:500val}, reveals that the predictions are not well calibrated. 
The predictions do not accurately reflect disease severity on a more granular scale than the labels used for model training. 
\\
Siamese networks are not trained to predict a specific value, unlike the other models, but rather to detect whether two images stem from the same or different classes. \cite{Koch2015SiameseRecognition}
Therefore, they can pick up subtle differences in disease severity. \cite{Li2020SiameseImaging}
Here, we obtained predictions comparing the input image of interest to a pool of anchor images that are typical representations of the class corresponding to the lowest label score.
While the predicted difference between the anchor images and the target images resulted in accurate ordinal predictions (Figure \ref{fig:rank_vs_pred}, it was not well calibrated to the underlying continuous variable, particularly at the extremes. 

\paragraph{MC dropout improves prediction of continuous variables}
Through the use of MC dropout, all four model types evaluated showed an improvement in the quality of the continuous scores as reflected in significantly higher Spearman correlation coefficients and lower MSE (see Table \ref{tab:metric-table}). 
MC multi-class classification networks were consistently among the highest performing models for all tasks and datasets evaluated, making them the top-performing models in our study.
\\
MC dropout presents a simple way to obtain meaningful continuous predictions from models trained using ordinal labels without sacrificing and, in some cases, even significantly improving predictive performance (see Table \ref{tab:metric-table}). 
However, MC dropout comes at a higher computational cost as inference requires multiple passes of the same input image to obtain the final prediction.
If the additional computational burden is a concern, ordinal classification or regression are alternatives to conventional multi-class classification models that are easy to train and provide decent continuous predictions without the use of MC dropout.


\paragraph{Limitations}
There are some limitations to this study. 
First, we treated the available ordinal labels as ground truth. 
For all three image analysis tasks analyzed here, high inter-rater variability, particularly around the decision boundaries between severity classes, have been reported. \cite{Campbell2016PlusVariability, Kalpathy-Cramer2016PlusAnalysis, Reijman2004ValidityApproach, RedondoInter-andMammograms}
It would be desirable for future work to explore the influence of noisy and biased ordinal ratings for the task of learning and predicting a continuous variable. 
Second, due to the latent nature of the variable of interest, for most of our analysis, we had to rely on proxy variables such as rankings and more granular expert disease severity ratings. 
Lastly, MC dropout predictions were based on 50 samples, an empirically chosen value based on common practices and our own experience. 

\section{Conclusion}\label{conclusion}
In this work, we present a generalizable framework to predict meaningful continuous scores while only using discrete ordinal labels for model development. Our findings are particularly relevant to disease severity prediction tasks as the available labels are usually coarse and ordinal, but continuous disease severity predictions could provide crucial information that allows for earlier detection of deterioration and more personalized treatment planning.

\backmatter





\bmhead{Acknowledgments}

The authors would like to thank Laura Coombs from the American College of Radiology for providing the DMIST dataset and the Volpara Density scores. 

\section*{Declarations}

\subsection{Funding}
J.P.C and S.O. are funded by the National Institutes of Health (Bethesda, MD) [R01 HD107493], an investigator-initiated grant from Genentech (San Francisco, CA) [R21 EY031883], and by unrestricted departmental funding and a Career Development Award (J.P.C.) from Research to Prevent Blindness (New York, NY) [P30 EY10572].
M.F.C. previously received grant funding from the National Institutes of Health (Bethesda, MD), and National Science Foundation (Arlington, VA).
J.P.C, S.O., M.F.C., J.K-C., and K.H. are supported by research funding from Genentech (San Francisco, CA)[R21 EY031883]. 
J.K-C. and K.V.H. are supported by funding from the National Institutes of Health (Bethesda, MD) [R01 HD107493] and National Cancer Institute (Bethesda, MD) [U01CA242879].
A.L. has a scholarship from Mitacs [IT24359], NSERC, and “Fondation et Alumni de Polytechnique Montréal”.

\subsection{Conflict of interest}
M.F.C. is an unpaid member of the scientific advisory board for Clarity Medical Systems (Pleasanton, CA), was previously a Consultant for Novartis (Basel, Switzerland) and was previously an equity owner at InTeleretina, LLC (Honolulu, HI). 
Dr. Campbell was a consultant to Boston AI Lab (Boston, MA), and is an equity owner of Siloam Vision.J.K-C. is a consultant/advisory board member for Infotech, Soft.
\\
The other authors declare no competing financial or non-financial interests.


\subsection{Code availability}
The code used to train the models can be found at \url{https://github.com/andreanne-lemay/gray_zone_assessment}.

\subsection{Data availability}
Access to the MOST dataset for knee osteoarthritis can be requested through the NIA Aging Research Biobank \url{https://agingresearchbiobank.nia.nih.gov/}. The breast density, and ROP datasets are not publicly accessible due to patient privacy restrictions.

\subsection{Author's contributions}

Study concept and design: K.H., A.L., J.K.-C., J.P.C. Data collection: J.P.C., S.O., and J.K.-C. Data analysis and interpretation: all authors. Drafting of the manuscript: A.L., K.H. Critical revision of the manuscript for important intellectual content and final approval: all authors. Supervision: J.K.-C., J.P.C.









\begin{appendices}
\section{Dataset label distributions \label{app:data}}
List of label distributions for each dataset.
\paragraph{Retinopathy of prematurity}
Dataset size: 5511 images
\begin{itemize}
\item Normal: 4535 images ($82{\cdot }3\%$)
\item Pre-plus disease: 804 images ($14{\cdot }6\%$)
\item Plus disease: 172 images ($3{\cdot }1\%$)
\end{itemize}

\paragraph{Knee osteoarthritis (OA)}
Dataset size: 14173 images
\begin{itemize}
\item No OA (KL 0): 5793 images ($40.9{\cdot }\%$) 
\item Doubtful OA (KL 1): 2156 images ($15{\cdot }2\%$) 
\item Mild OA (KL 2): 2355 images ($16{\cdot }6\%$)
\item Moderate OA (KL 3): 2604 images ($18{\cdot }4\%$)
\item Severe OA (KL 4): 1265 images ($8{\cdot }9\%$)
\end{itemize}

\paragraph{Breast density}
Dataset size: 108230 images 
\begin{itemize}
\item Fatty: 12428 images ($11{\cdot }5\%$) 
\item Scattered: 47909 images ($44{\cdot }2\%$)
\item Heterogeneously dense: 41325 images ($38{\cdot }2\%)$ 
\item Dense: 6568 images ($6{\cdot }1\%$)
\end{itemize}

\section{Model training parameters}\label{app:training}
\paragraph{ROP}
ROP models had a ResNet18 architecture and were trained with a batch size of 24, a learning rate of 1e-4 for 25 epochs, and the best model was selected using the highest accuracy on the validation set. Balanced class sampling mitigated the class imbalance during training. Data augmentation consisted of random rotation of $\pm$ 15 degrees with a probability of 0.5, random flips with a probability of 0.5, and random zooms of 0.9 to 1.1 with a probability of 0.5.

\paragraph{Knee osteoarthritis}
ResNet50 architecture was selected for the knee osteoarthritis model and was trained with the following parameters: batch size of 16, learning rate of 5e-6, 75 epochs. The final model was chosen based on the best loss value on the validation set. The data sampler used balanced weights during training to help with data imbalance. Images were randomly rotated of $\pm$ 15 degrees with a probability of 0.5 and randomly flipped with a probability of 0.5 as data augmentation.

\paragraph{Breast density}
Breast density models were trained with a ResNet50 architecture for 75 epochs by batches of 8 with a learning rate of 5e-5. The best model was selected using the best loss score on the validation set. The same data augmentation as the knee osteoarthritis model was applied for breast density.
\section{Predicted rank vs. ground truth rank}\label{sec:app_rank}
\begin{figure}[H]
     \centering
     \begin{subfigure}[b]{\textwidth}
         \centering
         \includegraphics[width=0.8\textwidth]{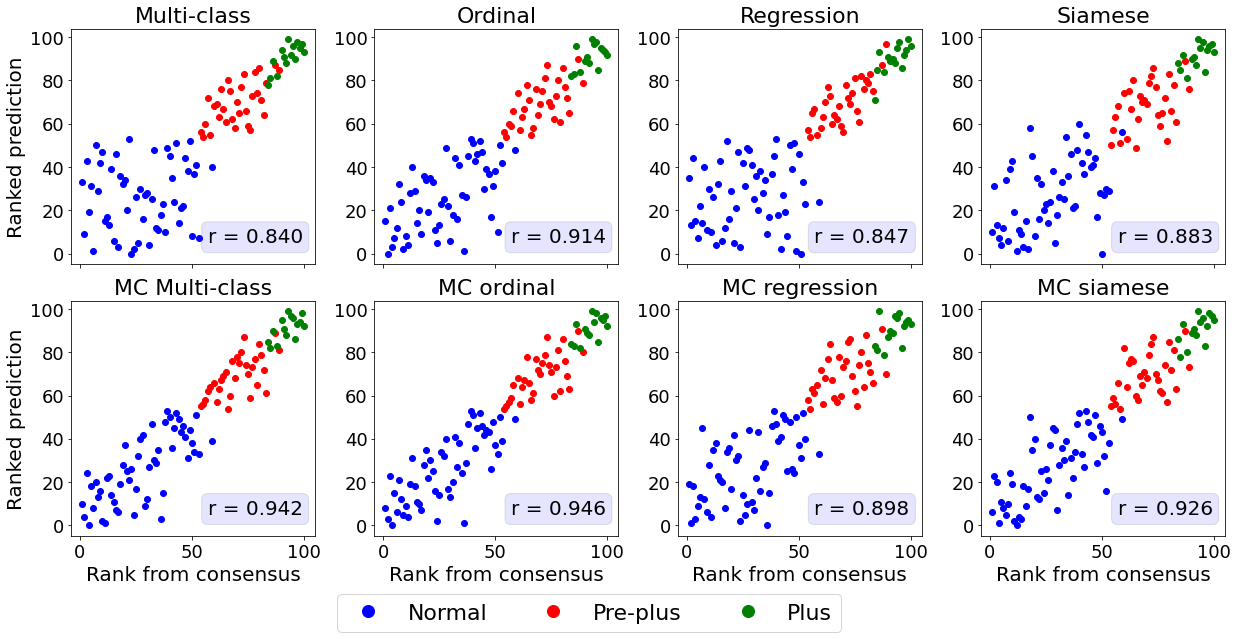}
         \caption{ROP}
         \label{fig:rop_rank}
     \end{subfigure}
     \hfill
     \begin{subfigure}[b]{\textwidth}
         \centering
         \includegraphics[width=0.8\textwidth]{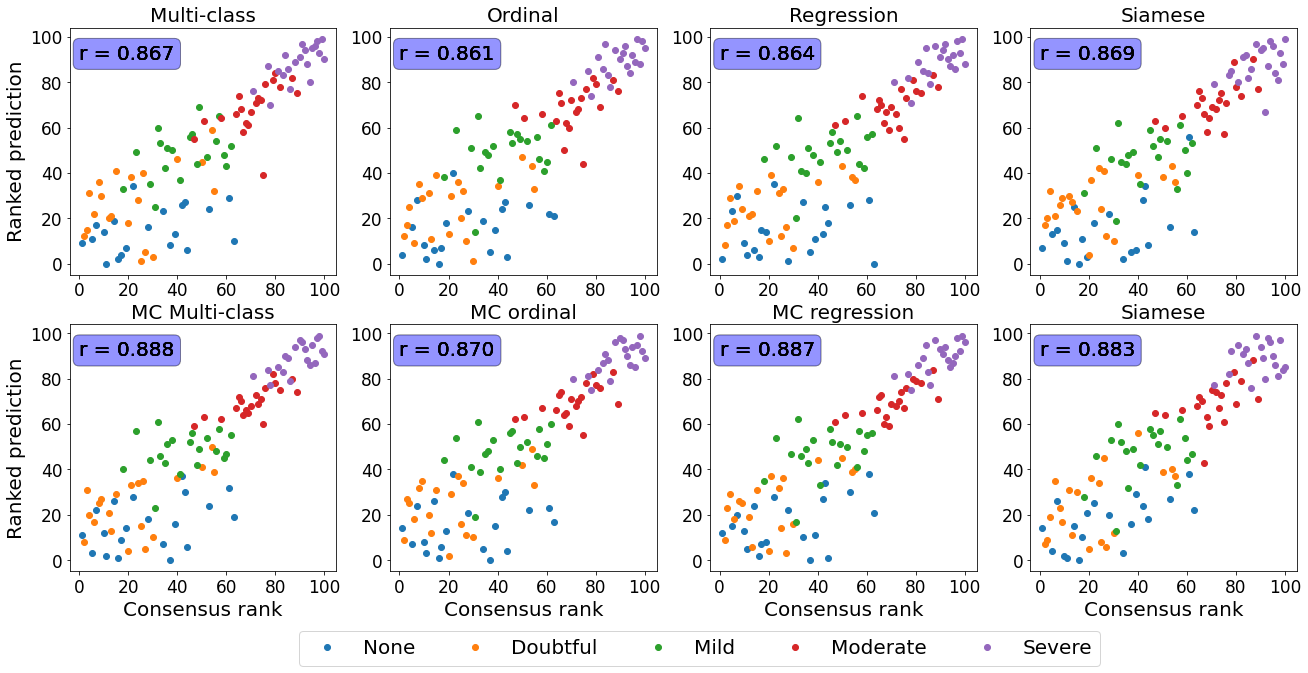}
         \caption{Knee osteoarthritis}
         \label{fig:knee_rank}
     \end{subfigure}
     \hfill
     \begin{subfigure}[b]{\textwidth}
         \centering
         \includegraphics[width=0.8\textwidth]{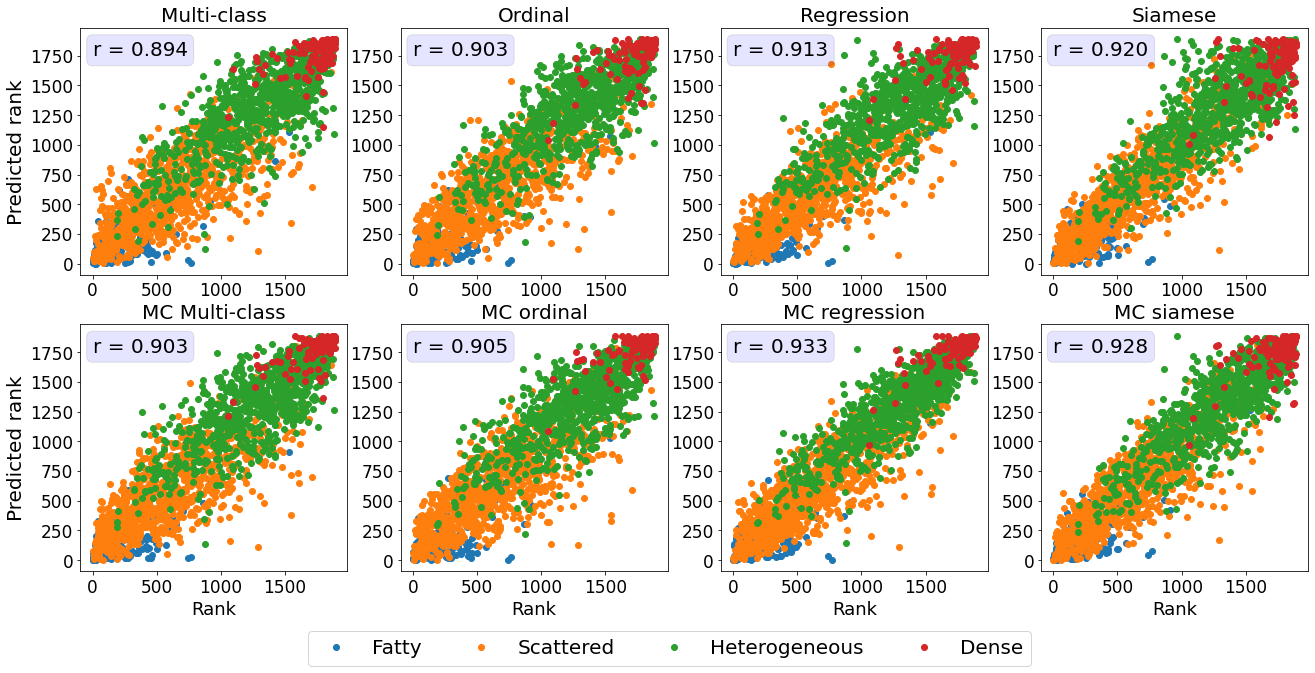}
         \caption{Breast density}
         \label{fig:mammo_rank}
     \end{subfigure}
        \caption{\textbf{Correspondence between model predicted rank and true severity rank.} For each model the Pearson correlation coefficient ($r$) is displayed and indicate the strength of the linear correlation where 1 is a perfectly positive linear correlation and -1 a perfectly negative linear correlation.}
        \label{fig:rank_vs_rank}
\end{figure}

Figure \ref{fig:rank_vs_rank} contains the same data from Figure \ref{fig:rank_vs_pred} presented in Section \ref{sec:rank}. The predicted scores were ordered to determine a rank and were plotted against the expert's ranks. The MSE displayed in Table \ref{tab:metric-table} was computed on these two variables. Since a linear correlation is expected on the rank-to-rank analysis, the Pearson coefficient was used. 

\section{Pair-wise statistical comparisons}\label{sec:stats}
Only the metrics showing no statistical differences between two metrics were included (only some metrics from Table \ref{tab:metric-table}). If no figure for a specific metric and dataset is present, it means all the pair-wise comparisons showed a statistical difference. All metrics presented in Table \ref{tab:metric-table}, Figure \ref{fig:500val}, Figure \ref{fig:longitudinal}, and Figure \ref{fig:nepal} were analysed.

\begin{figure}[H]
     \centering

    \begin{subfigure}[b]{\textwidth}
         \centering
         \includegraphics[width=\textwidth]{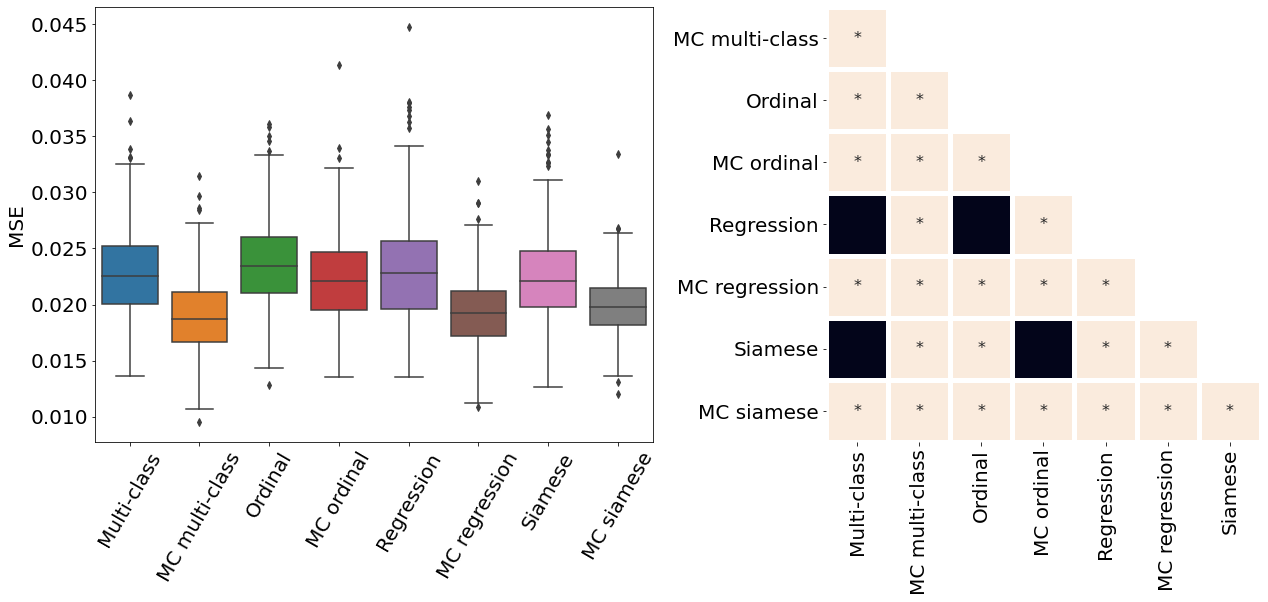}
         \caption{MSE - Knee osteoarthritis}
         \label{fig:rop100auc}
     \end{subfigure}
     \hfill
     \begin{subfigure}[b]{\textwidth}
         \centering
         \includegraphics[width=\textwidth]{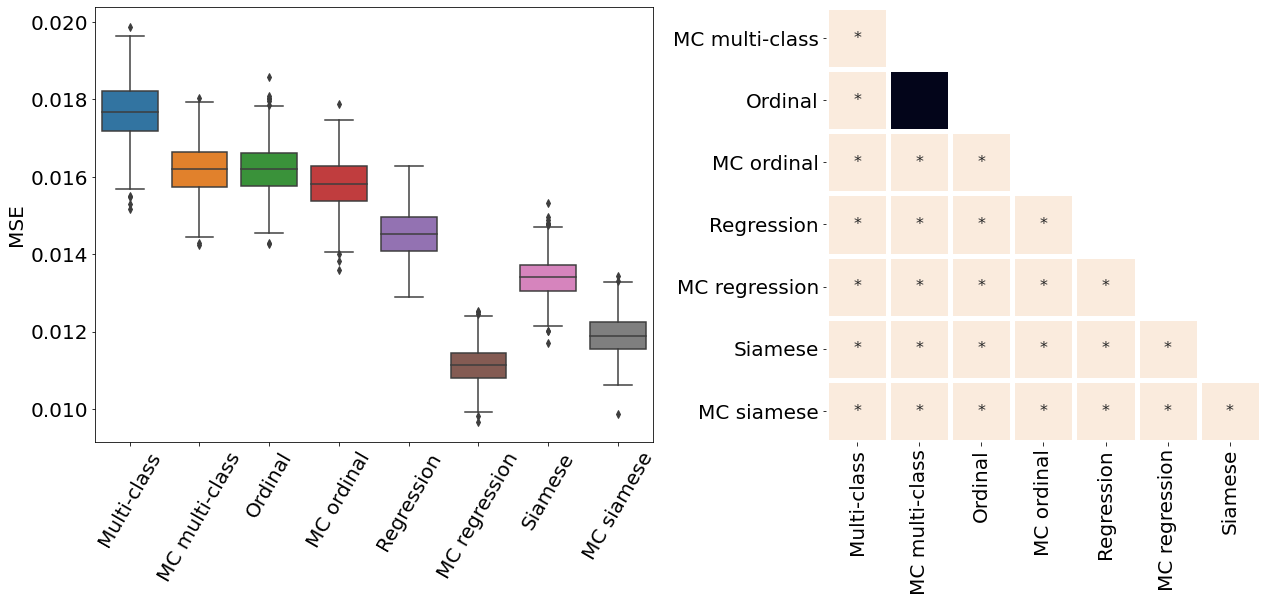}
         \caption{MSE - Breast density}
         \label{fig:mammomse}
     \end{subfigure}

\end{figure}

\begin{figure}[H]
\ContinuedFloat 
     \centering
     \begin{subfigure}[b]{\textwidth}
         \centering
         \includegraphics[width=\textwidth]{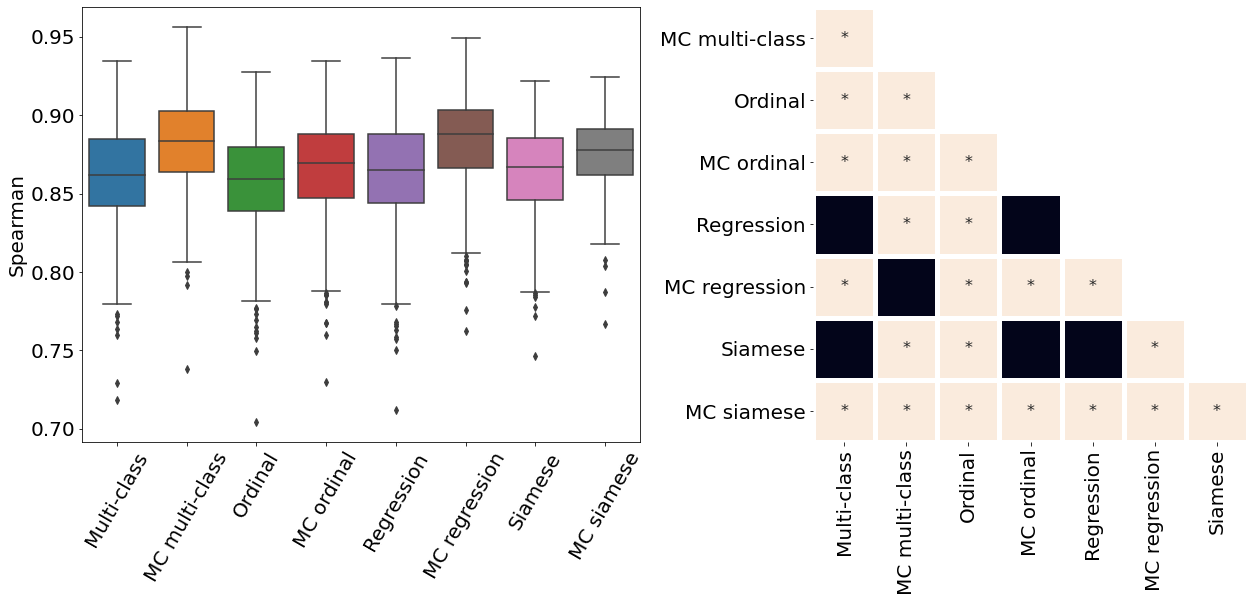}
         \caption{Spearman - Knee osteoarthritis}
         \label{fig:kneespearman}
     \end{subfigure}
     \hfill
     \begin{subfigure}[b]{\textwidth}
         \centering
         \includegraphics[width=\textwidth]{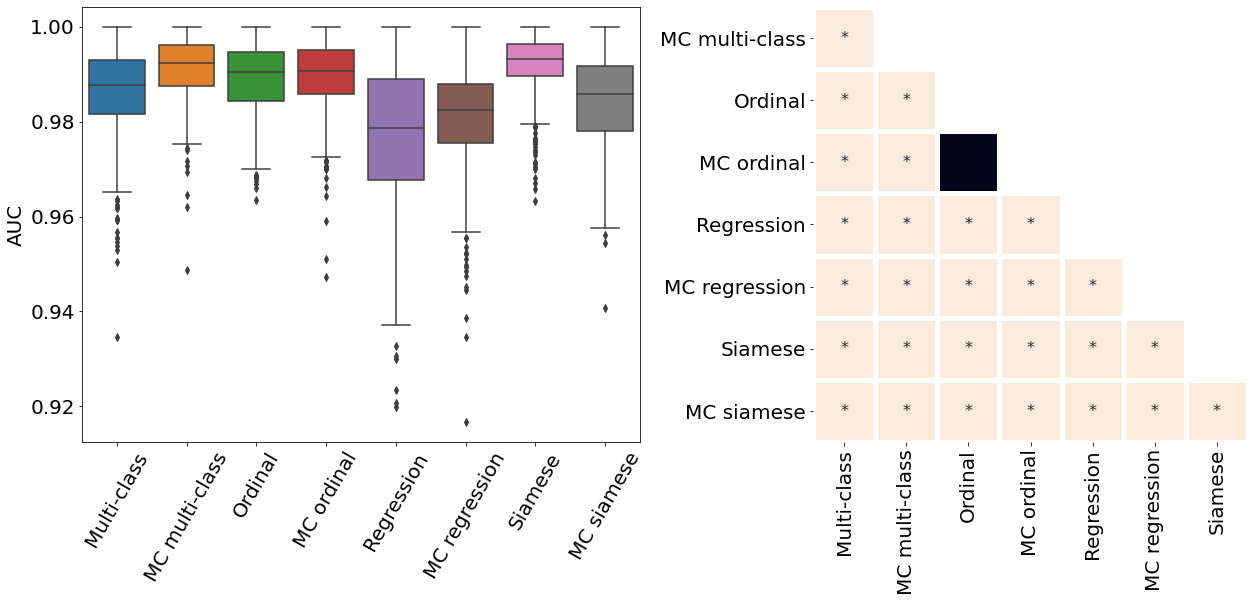}
         \caption{AUC - ROP (100 ranked cases)}
         \label{fig:rop100auc}
     \end{subfigure}
     \hfill
    \begin{subfigure}[b]{\textwidth}
         \centering
         \includegraphics[width=\textwidth]{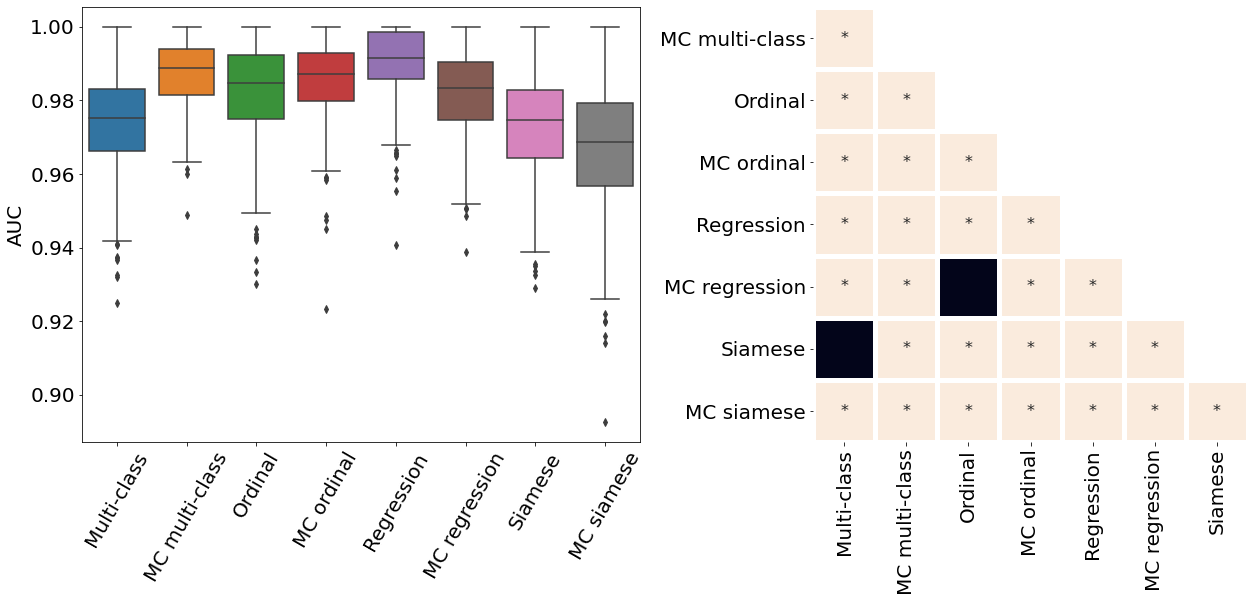}
         \caption{AUC - Knee osteoarthritis}
         \label{fig:mammoauc}
     \end{subfigure}
\end{figure}

\begin{figure}[H]
\ContinuedFloat 
     \centering
     \begin{subfigure}[b]{\textwidth}
         \centering
         \includegraphics[width=\textwidth]{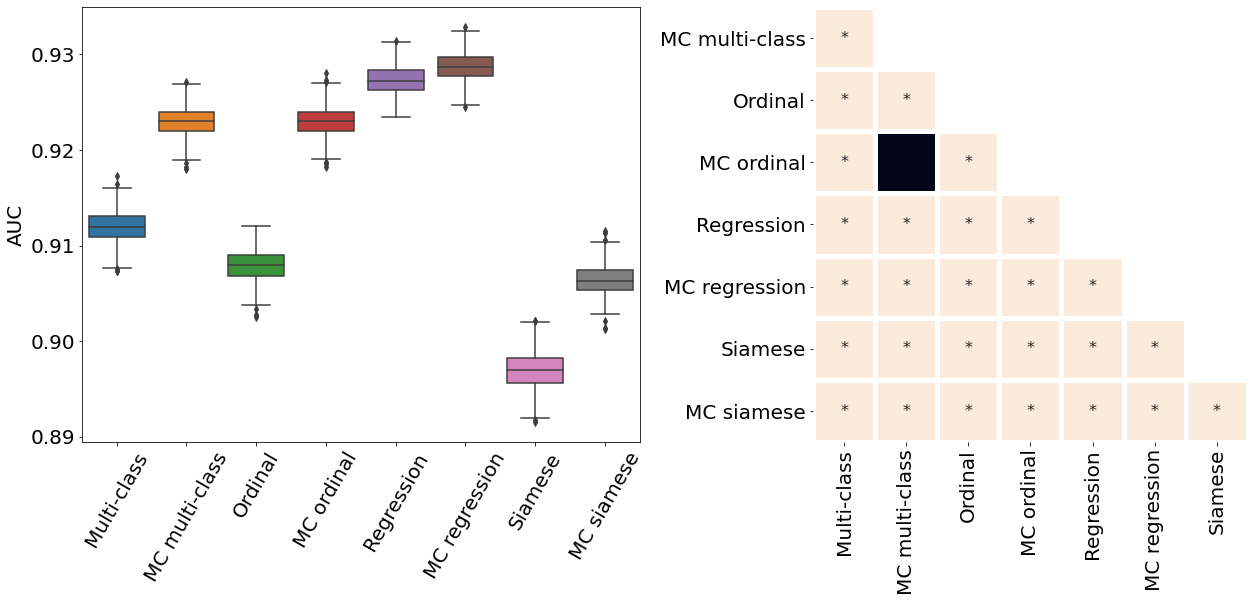}
         \caption{AUC- Breast Density}
         \label{fig:mammoauc}
     \end{subfigure}

    \caption{\textbf{Pair-wise statistical comparisons for MSE, Spearman, and AUC metrics (metric - dataset).} For each metric on a given test set, each pairs of models (MC and non-MC multi-class, ordinal, regression, Siamese) was compared. The box plots on the left side displays the value range obtained through 500 bootstraps. The grid on the right side includes the 28 pair-wise comparisons. * means that a statistical difference ($p-value < 0.05$ on a two-sided t-test) was reached while a black square indicates no statistical differences. Only metrics where at least one pair had no statistical difference was presented. }
        \label{fig:stats}
\end{figure}

\section{ROP - Score on out-of-distribution dataset}
\begin{figure}[H]
  \centering
    \includegraphics[width=1.0\textwidth]{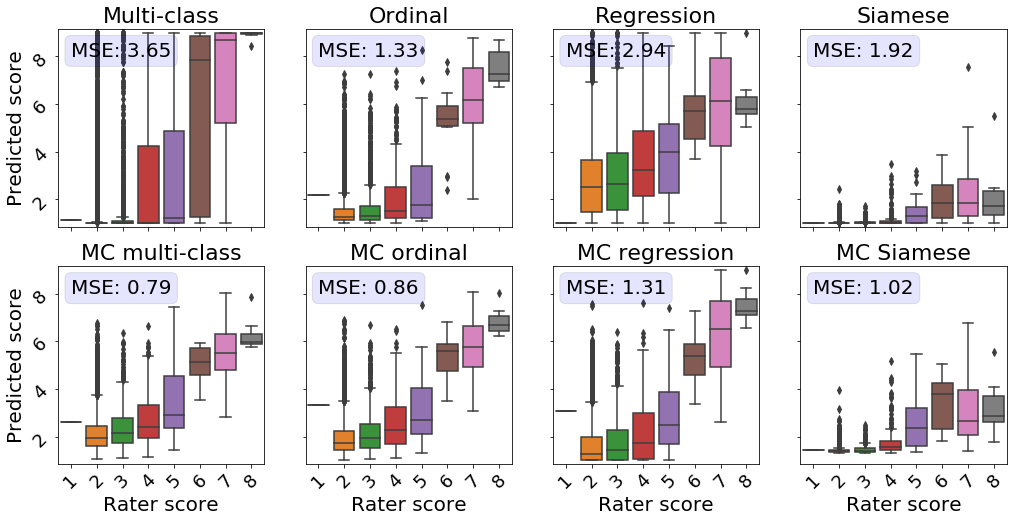}

  \caption{\textbf{Correspondence between predicted and consensus ROP severity on an out-of-distribution test set.} The rater score is obtained from a single rater. The predicted scores from multi-class, ordinal, and regression models that were trained to predict values from 0 to 2 were scaled and shifted to match the 1 to 9 range ($score_{rescaled} = score_{model} \times 2 + 1$). Siamese networks predict values from 0 to infinity and is not fully bounded. The Siamese scores were hence only shifted by 1 ($score_{rescaled} = score_{Siamese} + 1$). All MSE measurements reported in this figure are statistically different $(p-value < 0.05)$.}
  \label{fig:nepal}
\end{figure}

The ROP models were further tested on a dataset from a different population and acquired at different centers from the training dataset. Figure \ref{fig:nepal} illustrates the correspondence between the predicted and rater scores for this out-of-distribution dataset. Similar to the in-distribution test set (see Figure \ref{fig:500val}), all the MC models had a better MSE compared with the non-MC corresponding models and MC multi-class and MC ordinal are the best performing models. The multi-class and regression models for most severity scores predicted a wide range of values, often from 1 to 9 which could lead to medical errors. The miscalibration of Siamese models is especially noticeable in Figure \ref{fig:nepal} as visually, the predicted and rater score do not match for high severity values. This out-of-distribution dataset contains only a few plus and pre-plus images, i.e., only 328 plus and pre-plus cases compared to 7565 normal cases.
Driven by a large number of outliers particularly within images with the lower disease severity ratings (normal cases), the MSE is particularly high for the conventional multi-class, ordinal classification, and regression models.
The low number of images with higher disease severity scores also explains why the MSE is not extremely high even though the Siamese networks are visually miscalibrated.

\end{appendices}


\bibliography{sn-bibliography, references_kathi}


\end{document}